# AMINO ACID INTERACTION NETWORK PREDICTION USING MULTI-OBJECTIVE OPTIMIZATION

Md. Shiplu Hawlader[1] and Saifuddin Md. Tareeq[2]

[1]Department of Computer Science & Engineering, University of Asia Pacific, Dhaka, Bangladesh
shiplu.cse@uap-bd.edu

[2]Department of Computer Science & Engineering, University of Dhaka, Dhaka, Bangladesh
smtareeq@cse.univdhaka.edu

## ABSTRACT

*Protein can be represented by amino acid interaction network. This network is a graph whose vertices are the proteins amino acids and whose edges are the interactions between them. This interaction network is the first step of proteins three-dimensional structure prediction. In this paper we present a multi-objective evolutionary algorithm for interaction prediction and ant colony probabilistic optimization algorithm is used to confirm the interaction.*

## 1. INTRODUCTION

Proteins are biological macromolecules performing a vast array of cellular functions within living organisms. The roles played by proteins are complex and varied from cell to cell and protein to protein. The best known role of proteins in a cell is performed as enzymes, which catalyze chemical reaction and increase speed several orders of magnitude, with a remarkable specificity. And the speed of multiple chemical reactions is essential to the organism survival like DNA replication, DNA repair and transcription. Proteins are storage house of a cell and transports small molecules or ions, control the passages of molecules through the cell membranes, and so forth. Hormone, another kind of protein, transmits information and allow the regulation of complex cellular processes.

Genome sequencing projects generate an ever increasing number of protein sequences. For example, the Human Genome Project has identified over 30,000 genes [1] which may encode about 100,000 proteins. One of the first tasks when annotating a new genome is to assign functions to the proteins produced by the genes. To fully understand the biological functions of proteins, the knowledge of their structure is essential.

Proteins are amino acids chain bonded together in peptide bonds, and naturally adopt a native compact three-dimensional form. The process of forming three-dimensional structure of a protein is called protein folding and this is not fully understood yet in System Biologoy. The process is a result of interaction between amino acids which form chemical bond to make protein structure.

In this paper, we proposed a new algorithm to predict a interaction network of amino acids using two new emerging optimization techniques, multi-objective optimization based on evolutionary clustering and ant colony optimization.

## 2. AMINO ACID INTERACTION NETWORK

Amino acids are the building blocks of proteins. Protein a sequences of amino acids linked by peptide bond. Each amino acid has the same fundamental structure, differing only in the side-chain, designated the R-group. The carbon atom to which the amino group, carboxyl group, and side chain (R-group) are attached is the alpha carbon (*Cα*). The alpha carbon is the common reference point for coordinates of an amino acid structure. Among the 20 amino acids some are acidic, some are basic, some are polar, some non-polar. To make a protein, these amino acids are joined together in a polypeptide chain through the formation of a peptide bond. The structure, function and general properties of a protein are all determined by the sequence of amino acids that makes up the primary sequence. The primary structure of a protein is the linear sequence of its amino acid structural units and it is a part of whole protein structure. The two torsion angles of the polypeptide chain, also called Ramachandran angles, describe the rotations of the polypeptide backbone around the bonds between *N – Cα* (called Phi angle, φ) and *Cα – C* (called Psi angle, ψ). Torsion angle is one of the most important parameter of protein structure and controls the protein folding. For each type of the secondary structure elements there is a characteristic range of torsion angle values, which can clearly be seen on the Ramachnadran plot [2].

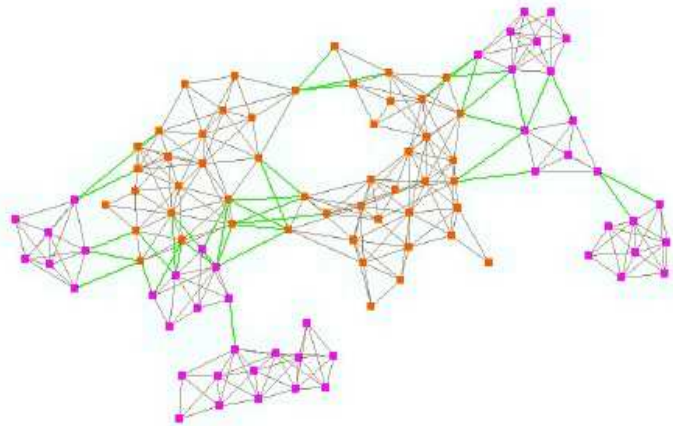

Figure 1: SSE-IN of 1DTP protein. Green edges are to be
predicted by ant colony algorithm

Another important property of protein is hydrophobicity. Proteins tertiary structure's core are hydrophobic and the amino acids inside core part do not interact much as like their counterpart hydrophilic, those made the outer side of the protein structure. Many systems, both natural and artificial, can be represented by networks, that is by site or vertices connected by link or edges. Protein also be represented as a network of amino acid whose edges are the interactions or the protein functions between amino acids. The 3D structure of a protein is determined by the coordinates of its atoms. This information is available in Protein Data Bank (PDB) [3], which regroups all experimentally solved protein structures. Using the coordinates of two atoms, one can compute the distance between them. We define the distance between two amino acids as the distance between their *Cα* atoms. Considering the *Cα* atom as a center of the amino acid is an approximation, but it works well enough for our purposes. Let us denote by *N* the number of amino acids in the protein. A contact map matrix is an *N X N, 0 - 1* matrix, whose element *(i, j)* is 1 if there is a contact between amino acids *i* and *j* and 0 otherwise. It provides useful information about the protein. For example, the secondary structure elements can be identified using this matrix. Indeed, α - helices spread along the main diagonal, while β - sheets appear as bands parallel or perpendicular to the main diagonal [4]. There are different ways to define the contact between two amino acids. In [5], the notion is based on spacial proximity, so that the contact map can consider non–covalent interactions. Gaci et al. in [5] says that two amino acids are in contact iff the distance between them is below a given threshold. A commonly used threshold is 7 Å. Consider a contact map graph with N vertices (each vertex corresponds to an amino acid) and the contact map matrix as incidence matrix. It is called contact map graph. The contact map graph is an abstract description of the protein structure taking into account only the interactions between the amino acids. Now let us consider the sub-graph induced by the set of amino acids participating in SSE, where SSE is secondary structure element of protein like alpha helix, beta sheet etc. We call this graph SSE interaction network (SSE - IN). The reason of ignoring the amino acids not participating in SSE is simple. Evolution tends to preserve the structural core of proteins composed from SSE. In the other hand, the loops (regions between SSE) are not so important to the structure and hence, are subject to more mutations. That is why homologous proteins tend to have relatively preserved structural cores and variable loop regions. Thus, the structure determining interactions are those between amino acids belonging to the same SSE on local level and between different SSEs on global level. In [6] and [7] the authors rely on similar models of amino acid interaction networks to study some of their properties, in particular concerning the role played by certain nodes or comparing the graph to general interaction networks models. Thanks to this point of view the protein folding problem can be tackled by graph theory approaches.

Gaci et al. in [8], has described the topological properties of a network and compared them with some *All alpha* and *beta* to prove that a protein can be treat as a network of amino acids. According to the diameter value, average mean degree and clustering coefficient shown in the experiment in [8], we can say a protein is a network of amino acids.

## 3. PREDICT AMINO ACID INTERACTION NETWORK

We can define the problem as prediction of a graph *G* consist of *N* vertices *V* and *E* edges. If two amino acids interact with each other in protein we mention it as an edge *(u, v)* ∈ *E, u* ∈ *V, v* ∈ *V* of the graph. A SSE-IN is a highly dense sub-graph $G_{SEE-IN}$ with edge set $E_{SEE-IN}$. Probability of the edge *(u, v)* ∈ $E_{SEE-INA}$, *u* ∈ $V_{SSE-INA}$, *v* ∈ $V_{SSE-INA}$ is very high and probability of the edge *(u, v)* ∉ $E_{SEE-INA}$, *u* ∈ $V_{SSE-INA}$, *v* ε $V_{SSE-INA}$ is very low where $V_{SSE-INA}$ and $V_{SSE-INB}$ are respectively the vertex set of SSE-IN A and SSE-IN B. SCOP and CATH are the two databases generally accepted as the two main authorities in the world of fold classification. According to SCOP there are 1393 different folds. To predict the network we have to solve three problems, as i) find a associate SCOP protein family from the given protein sequence ii) predict a network of amino acid secondary structure element (SSE) from the known SCOP protein family and iii) Predict interactions between amino acids in the network, including internal edges of SSE-IN and external edges.

We are going to avoid the description the first problem, because it can be solve using a good sequence alignment algorithm like BLAST as discussed in [8]. In this paper we are going to solve the second and third problem individually with multi-objective optimization using genetic algorithm and ant colony optimization respectively.

Gaci et al. in [9], described a solution to the prediction of amino acid interaction network. He used a genetic algorithm with single objective as the distance between to amino acid in protein atom. But it is very difficult to define real world problems like amino acid interaction problem in terms of a single objective. A multi-objective optimization problem deals with more than one objective functions that are to be minimized or maximized. These objectives can be conflicting, subject to certain constraints and often lead to choosing the best trade-off among them. As we have described before, the interaction between amino acids in protein depends not only distance between two amino acids but also the torsion angles and hydrophobic property of the amino acid. So to get more accurate interaction network of amino acid we have to consider it is as a multi-objective problem rather than single objective.

## 4. ALGORITHM

As we have mentioned before, we can solve the amino acid interaction network prediction problem as well as the protein folding problem using two new and emerging algorithms. The multi-objective optimization algorithm will predict structural motifs of a protein and will give a network or graph of secondary structural element (SSE) of the protein. On the other hand, the ant colony optimization (ACO) algorithm will find the interactions between amino acids including the intra-SSE-IN and inter-SSE-IN interactions. In our algorithm we have considered a folded protein in the PDB as an unknown sequence if it has no SCOP v1.73 family classification. According to [8], we can associate the most compatible and best fit structural family based on topological criteria like average diameter, average mean distance etc.

### 4.1. Prediction of SSE interaction network using Multi-objective Optimization

There are several ways to solve multi-objective optimization problem. In this research we have decided to use Genetic Algorithm (GA) as multi-objective optimization. The GA has to predict the adjacency matrix of unknown sequence when it is represented by chromosome.

In this paper we proposed a evolutionary clustering algorithm to predict the SSE-IN, which is a modified algorithm of the second version of strength pareto evolutionary algorithm (SPEA2) in [10]. SPEA2 preserve better solutions than NSGA-II [11] and its diversity mechanism is better than the others, this is the reason to choose SPEA2 to implement the evolutionary clustering algorithm.

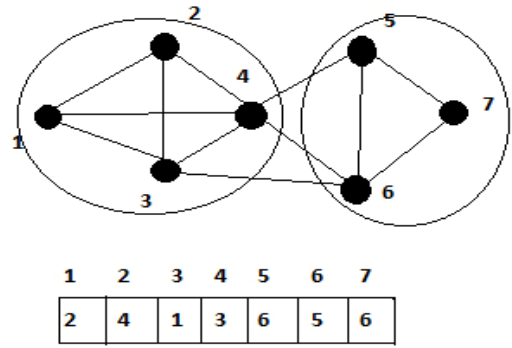

Figure 2: Network of 7 nodes clustered into 1,2,3,4 and 5,6,7
and their genetic representation

As proposed in [12], we are using a local-based adjacency representation. In this representation an individual of the population consist of *N* genes $g_1,...,g_N$, where *N* is the number of nodes. Each gene can hold allele value in the range *1,...,N*. Genes and alleles represents nodes in the graph *G = (V,E)* modelling a network *N*. A value *j* assigned in *i-th* gene interpreted as a link between node *i* and *j* and in clustering node *i* and *j* will be in the same cluster as in Figure 2. In decoding step all the components are identified and nodes participating in the same component are assigned to the same cluster.

**Alogorith1** Multiobjective genetic algorithm to predict SSE interaction algorithm

1: **Input:** A protein sequence, T = total time steps, $N_E$ = Archive size, $N_P$ = Population size
2: **Output:** A predicted incident matrix M and clustering for each network $N^i$ of $N$
3: Use BLAST to find a associate protein family of the given sequence from PDB
4: Generate initial cluster $CR_1 = \{C_1^1, \ldots, C_k^1\}$ of the network $N^1$ with number of vertex equal to number of SSE of the associate protein family
5: **for** t = 2 to T **do**
6:     Create initial population of random individual $P_0$ and set $E_0 = \emptyset$, $i = 0$
7:     **Loop**
8:         Decode each individual of $P_i \cup E_i$
9:         Evaluate each individual of $P_i \cup E_i$ to find rank and density value using equation 1 and 2
10:         Assign fitness value to each individual, as the sum of rank and inverse of density value
11:         Copy all no dominating solution to $E_{i+1}$
12:         **if** $|E_{i+1}| > N_E$ **then**
13:             truncate $|E_{i+1}| - N_E$ solutions according to topological property
14:         **Else**
15:             copy best $N_E - |E_{i+1}|$ dominated solution according to their fitness value and topological property
16:         **end if**
17:         **if** stopping criteria does not satisfies **then**
18:             **return** non-dominated solution in $|E_{i+1}|$
19:         **else**
20:             Select some individual form— $E_{i+1}$ for mating pool as parents using binary tournament with replacement
21:             Apply crossover and mutation operators to the mating pool to the mating pool to create $N_p$ offspring solution and copy to $P_{i+1}$
22:             i := i+1
23:         **end if**
24:     **end loop**
25:     From the returned solution in E take the best cluster according to the highest modularity value
26: **end for**

It takes a dynamic network $N = N_1, N_2, ... N_T$, the sequence of graphs $G = G_1, G_2, ... , G_T$ and the number of timestamps $T$ as input and gives a clustering of each network $N_i$ of $N$ as output.

In the amino acid interaction network, total number of gene is the number SSE in the associate protein family found from the first step and each SSE represents one gene or allele notably considering its size that is the number of amino acids which compose it, of the population. We represent a protein as an array of alleles. The position of an allele corresponds to the SSE position it represents in the sequence. At the same time, an incident matrix is associates for each genome.

For the first time-stamp of first input network there is no temporal relation with the previous network. The only objective function is snapshot quality or snapshot score. Thus we can apply any static clustering algorithm or trivial genetic algorithm to find the initial cluster. In this algorithm we used genetic algorithm to find the best cluster by maximizing the only objective function. As it is single objective algorithm we can find the single best cluster from this step.

As a first step in each time-stamp from $2^{nd}$ time-stamp to $T$, it creates a population of random individuals. Each individual is a vector of length equal to number of nodes in the graph $G^t$. Genetic variant operators will be applied on this population for a fixed number of pass.

Each individual of the population and archive is decoded into component graph. As each individual gene is working as an adjacency list, if a node in $x$ of graph is reachable from $y$ by maintaining the edges in the individual, then $x$ and $y$ is in same cluster of component.

Give each individual chromosome of the population and chromosome in archive a rank value. Smaller the rank value better it is as fitness value. Each non-dominated individual gets the rank 0. After removing the 0 ranked individuals, give the rank 1 to the next non-dominated individuals and so on. After giving each individuals a rank value, sort the individuals according to the ascending rank.

$$r(x) = \sum_{x \prec y} s(y) \qquad (1)$$

There could be many individuals of in same area of solution space or objective space. If we take all these solution into account, we could loss diversity in the population. To remain the population diverse, we are using distance of *k-th* nearest neighbour. The fitness value of each individual is the sum of its non-dominated rank and the inverse of the distance of *k-th* nearest neighbours distance. More the distance between solutions, better the fitness functions value.

$$m(x) = (\sigma_x^k + 1)^{-1} \qquad (2)$$

where $\sigma_x^k$ is the distance between individual $x$ and its *k-th* nearest neighbour. To calculate the distance between chromosome, we have to take account the three objectives, atomic distance of amino acids, torsion angles and hydrophobicity.

After evaluating fitness values of each of the population and archive, the best individuals are selected as a new population. From the total individuals of population and archive population size individuals are selected as new population. From the rank 0 to the highest rank, all the individuals are added if number of population of this rank is not exceeding the current population size. If it is exceeding, then some individuals are truncated according to the value of each individuals.

Table 1: Example of uniform crossover

| Parent 1 | *4* | *3* | *2* | *2* | *6* | *5* | *6* |
|---|---|---|---|---|---|---|---|
| Parent 2 | 3 | 3 | 1 | 5 | 4 | 7 | 6 |
| Mask | 0 | 1 | 1 | 0 | 0 | 1 | 1 |
| Offspring | 4 | 3 | 1 | 2 | 6 | 7 | 6 |

After selecting the new population, a mating pool is created of pool size from the new population to apply the genetic variation operators. To choose the mating pool, binary tournament with replacement has been used in this algorithm. According to binary tournament, two individuals are randomly selected from the new population and the better fitness valued individual is chosen for the mating pool.

#### 4.1.1. Genetic Variation Operators

Genetic operators are used to create offspring from parent or mating pool. As other genetic algorithms, in this algorithm two widely used genetic variation operators have been used. These are crossover and mutation.

Crossover is the operator which is used to create offspring from two parents. The offspring bear the genes of each parent. As a genetic variation operator there is very high probability to crossover occurs other than mutation. In this algorithm we are using uniform crossover. A random bit vector of length of number of the node in the current graph is created. If *i-th* bit is 0 then the value of the *i-th* gene comes from the first parent otherwise it comes from the *i-th* gene of second parent. As each of the parents holding true adjacency information, the offspring will also hold it.

One of the most widely used variation operator in genetic algorithm, which perform the operation in a single individual is mutation. Though the probability of mutation is normally very low, but it is the best way to make small variation in the individual. To mutate and create a offspring, some position of the of the individuals are chosen randomly and changed to other values. But the value should be one of its neighbours in the current graph.

A topological operator is used to exclude incompatible population generated by the algorithm. We compute the diameter, the characteristic path length and the mean degree to evaluate the average topological properties of the family for the particular SSE number.

### 4.1. Ant Colony Optimization (ACO) to Predict Interactions

After predicting the SSE-IN network we have to identify the interactions involve between the amino acids in the folded protein. We have used an ant colony optimization (ACO) approach to select and predict the edges which link different SSE's, considering about the correction of the matrix of motifs previously predicted.

We have built a two steps algorithm as the hierarchical structure of the SSE-IN.

- In interaction, consider each pair of SSE's separately. This is the local step. We use an ant colony algorithm to identify the suitable interactions between amino acids belonging to these SSE's.

- A global ant colony algorithm is run to predict the interaction between amino acids from different SSE-IN.

### 4.2.1. Parameters for Interaction Network Prediction

To predict the interactions, firstly we have to know how many edges to be add in the network and which nodes we should consider in interactions. To find and evaluate these parameters, we incorporated the template proteins from the associate family.

We select some template proteins from the associate family whose SSE number is same as the sequence to predict the edge rate of the sequence and represent them as chromosome or array of alleles as in the multi-objective genetic algorithm. Thus, we build a comparative model to compute the edge ratio, which is used to fold the sequence SSE-IN.

We calculate the average chromosome from all the template proteins in associate protein family. Here we used the distance between two chromosomes as discussed in the previous section to compare the sequence with the average chromosome. We add up the distance allele by allele to obtain a distance between the sequence and the average family chromosome. After that, we calculate the cumulated size by adding up the chromosome cell values. If the distance is less than 20% of the sequence cumulated size and the average family chromosome then the sequence is closer to the template protein. Then we compute the average edge rate in the closer protein to add the initial edges in the disconnected network of the sequence. If we can't find a sequence closer to the template one, we add the sequence with the average family chromosome and start again the same procedure.

We do the same procedure to find the designation of the vertices, which vertices should interact with each other as they also use comparative model.

To define, which edges link two SSE's, we consider the following problem.

Let $X = x_1, x_2, ... x_n$ and $Y = y_1, y_2, ... y_m$ be two SSEs in interaction. We want to add $e$ edges among the $n \times m$ possible combinations. For $i \in [1, n]$ and $j \in [1, m]$ the probability to interact the amino acid $x_i$ with $y_j$, is correlated with the occurrence matrix of the predicted edges ratios, represented by $Q(x_i, y_j)$ and we can assume $s_{ij} \sim Q(x_i, y_j)$. To add approximately $e$ edges, we need

$$\sum_{i=1}^{n}\sum_{j=1}^{m} S_{ij} = e \qquad (3)$$

and

$$S_{ij} = \frac{eQ(x_i, y_j)}{\sum_{p=1}^{n}\sum_{p=1}^{m} Q(x_p, y_q)} \qquad (4)$$

### 4.2.2. Ant Colony Algorithm

The prediction of interaction network consists of two approaches, local and global algorithm.

### 4.2.2.1. Local Algorithm

The local algorithm is used to predict the suitable shortcut edges between pair of SSEs in the network. Thus, we differentiate each pair of SSEs which have connection and build a graph where each vertex of the first SSE is connected to each vertex of the other SSE. The connection or the edges are weighted ($S_{ij}$). Then we used an ant colony approach consists of an ant number equals to the number of vertices in two SSE. The ant system has to reinforce the suitable edges between the SSEs. We use these edges in the global algorithm which is described in the next section.

The local ant colony algorithm first creates *n* ants which is total number of vertices in the two SSEs related in the search. For an ant to be positioned we choose a random vertex of to SSE involved and place it. All the *n* ants are positioned this way and two ants can share same vertex. An ant in vertex *i* will choose the vertex *j* with probability $p_{ij}$, defined as follows:

$$p_{ij} = \frac{\tau_{ij}^{\alpha} \cdot s_{ij}^{\beta}}{\sum_{k \in V(i)} \tau_{ij}^{\alpha} \cdot s_{ij}^{\beta}} \qquad (5)$$

The weight $s_{ij}$ also called heuristic vector, calculated before. If the vertices *i* and *j* are in the same SSE, then the edge between these two vertices has weight equal to the average weight of the shortcut edges:

$$\overline{S} = \frac{1}{nm} \sum_{i=1}^{n} \sum_{j=1}^{m} s_{ij} \qquad (6)$$

After each move of an ant we update the pheromone value on the inter-SSE edges using the formula,

$$\tau_{ij} = (1-\rho)\tau_{ij} + n_{ij}\Delta\tau \qquad (7)$$

where $s_{ij}$ is the number of ants that moved on the edge *(i, j)* and $\Delta\tau$ is the quantity of pheromone dropped by each ant. As far as the edges belonging to the same SSE are concerned, we keep the pheromone rate on them equals to the average pheromone rate on the inter-SSE edges

$$\overline{\tau} = \frac{1}{nm} \sum_{i=1}^{n} \sum_{j=1}^{m} \tau_{ij} \qquad (8)$$

Ants are move inside an SSE randomly, described above, on the other hand if they decide to change the SSE they are guided by the edge weight and the weight is guided by the pheromone value. The algorithm stops after a predefined number of iteration or the maximum pheromone rate is *e* time bigger than the average pheromone rate on the edge. After the execution of the algorithm we keep the edges whose pheromone quantity exceeds a threshold $\lambda_{min}$.

**Algorithm 2** Local algorithm to find Inter-SSE edges

1:    **Input:** The predicted network from the multiobjective genetic algorithm

2:    **Output:** Predicted inter-SSE edges

3:    Create n ants, where n is the total number of nodes in process

4:    **while** stopping criteria does not meet **do**

5:       **for all** ant α **do**

6:          moveAnt(α)

7:       **end for**

8:       updatePheromone()

9:    **end while**

10:    selectEdges($\lambda_{min}$)

**Algorithm 3** Global algorithm to predict edges into SSEs

1:    **Input:** The network with predicted edges $E_s$ from local algorithm and $E_p$, number edges to predict

2:    **Output:** The network with total $E_p$ edges

3:    buildSSEIN($E_s$)

4:    create n ants

5:    **while** stopping criteria does not meet **do**

6:       **for all** ant a **do**

7:          moveAnt(α)

8:       **end for**

9:       updatePheromone()

10:    **end while**

11:    selectEdges($E_p$)

#### 4.2.2.2. Global Algorithm

After the local algorithm execution, we get the SSE-IN composed of these specific inter-SSE edges. The global algorithm will keep the number of edges exactly $E_p$, which was predicted before. As the local one, the global algorithm uses the ant colony approach with the number of vertices equal to the SSE-IN vertex number. The ants movements contribute to emerge the specific shortcut who's only a number $E_p$ is kept. We rank the shortcut edges as a function of the pheromone quantity to extract the $E_p$ final shortcuts. Finally, we measure the resulting SSE-INs by topological metrics to accept it or not.

We compute the diameter, the characteristic path length and the mean degree to evaluate the average topological properties of the family for a particular SSE number. Then, after we have built the sequence SSE-IN, we compare its topological properties with the template ones. We allow an error up to 20% to accept the built sequence SSE-IN. If the built SSE-IN is not compatible, it is rejected. We compare the predicted value, denoted $E_p$, with the real value, denoted $E_R$

$$AC = 1 - \frac{|E_R - E_p|}{E_p} \qquad (9)$$

where AC is the accuracy of the prediction.

## 5. PERFORMANCE ANALYSIS

In this paper we have discussed two algorithms to predict the interaction network of amino acid. We are going to analysis each algorithm independently.

### 5.1. Analysis of Genetic Algorithm as Multi-objective Optimization

In order to test the performance of proposed multi-objective genetic algorithm, we randomly pick three chromosomes from the final population and we compare their associated matrices to the sequence SSE-IN adjacency matrix. To evaluate the difference between two matrices, we use an error rate defined as the number of wrong elements divided by the size of the matrix. The dataset we use is composed of 698 proteins belonging to the *All alpha* class and 413 proteins belonging to the *All beta* class. A structural family has been associated to this dataset as in [8].

*All alpha* class has an average error rate of 14.6% and for the *All beta* class it is 13.1% and the maximum error rate shown in the experiment is 22.9%. Though, the error rate depends on other criteria like the three objectives described before but according to the result we can firmly assert that the error rate is depends on the number of initial population, more the number of initial population less the error rate. With sufficient number of individuals in the initial population we can ensure the genetic diversity as well as the improved SSE-IN prediction. When the number of initial population is at least 15, the error rate is always less than 10%.

As compared to the work in [8] we can claim better and improved error rate in this part of SSE-IN prediction algorithm.

### 5.2. Analysis of Ant Colony Optimization

We have experimented and tested this part of our proposed method according to the associated family protein because the probability of adding edge is determined by the family occurrence matrix. We have used the same dataset of sequences whose family has been deduced.

For each protein, we have done 150 simulations and when the topological properties are become compatible to the template properties of the protein we accepted the built SSE-IN. The results are shown in Table 2. The score is the percentage of correctly predicted shortcut edges between the sequence SSE-IN and the SSE-IN we have reconstructed [8]. In most cases, the number of edges to add were accurate according to the Figure 3. From this we can percept that, global interaction scores depends on the local algorithm lead for each pair of SSEs in contact. The plot, in Figure 3 confirms this tendency, if the local algorithm select at least 80% of the correct shortcut edges, the global intersection score stays better than the 80% and evolve around 85% for the All alpha class and 73% for the All beta class.

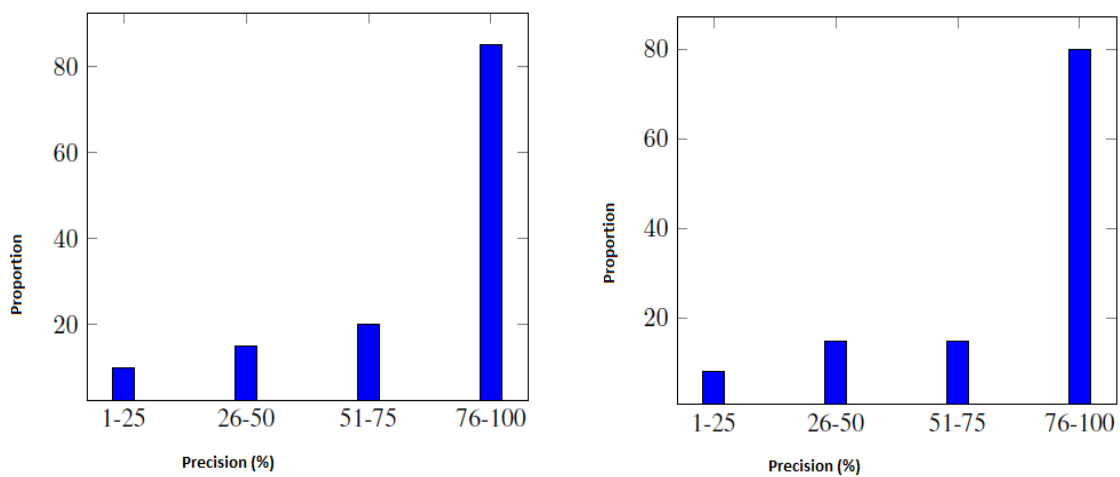

Figure 3: Precision in number of edges to be added in All Alpha (left) and All Beta (right).

After the discussion we can say that, though for the big protein of size more than 250 amino acids the average score decreases, but in an average the score remains for the global algorithm around 80%.

**5.3. Algorithm Complexity**

Our proposed algorithm is independent of specific time bound. Both the optimization algorithm used as multi-objective genetic algorithm and ant colony algorithm, is iteration based. We can stop the algorithm at any time. Though the result of the algorithm depends on the number of iteration but if we give sufficient amount of iteration it provides good result. In compare to other state of art algorithms, those uses exponential complexity algorithm, our is linear in terms of time and memory.

# 6. CONCLUSIONS

We have proposed an computational solution to an biological problem. We have described how we can formulate a biological problem like folding protein into optimization and graph theory problem. The formulation consists of finding the interactions between secondary structure element (SSE) network and interaction between amino acids of the protein. The first problem was solving by an multi-objective genetic algorithm and the second one solve by ant colony optimization approach.

As discussed before, we have given theoretical and statically proof that our proposed algorithm gives more accurate result in terms of accuracy and score to predict the amino acid interaction

Table 2: Folding a SSE-In by ant colony approach. The algorithm parameter values are : $\alpha = 25$, $\beta = 12$, $\rho = 0.7$, $\Delta\tau = 4000$, $e = 2$, $\lambda_{min} = 0.8$.

| Class | SCOP Family | Number of Proteins | Protein Size | Score | Average Deviation |
|---|---|---|---|---|---|
| All Alpha | 46688 | 17 | 27-46 | 83.973 | 3.277 |
| | 47472 | 10 | 98-125 | 73.973 | 12.635 |
| | 46457 | 25 | 129-135 | 76.125 | 7.849 |
| | 48112 | 11 | 194-200 | 69.234 | 14.008 |
| | 48507 | 18 | 203-214 | 66.826 | 5.504 |
| | 46457 | 16 | 241-281 | 63.281 | 17.025 |
| | 48507 | 20 | 387-422 | 62.072 | 9.304 |
| All Beta | 50629 | 6 | 54-66 | 79.635 | 2.892 |
| | 50813 | 11 | 90-111 | 74.006 | 4.428 |
| | 48725 | 24 | 120-124 | 80.881 | 7.775 |
| | 50629 | 13 | 124-128 | 76.379 | 9.361 |
| | 50875 | 14 | 133-224 | 77.959 | 10.67 |

network. Though it can be furnished further with improved data structure and parallel algorithms.

**Authors**

Md. Shiplu Hawlader, born 1987 in Dhaka, Bangladesh, graduated from the Faculty of Engineering of University of Dhaka in 2011 at the Computer Science and Engineering department. After the graduation he completed his M.Sc.degree at the same department. He is a lecturer of Department of Computer Science and Engineering, University of Asia Pacific.

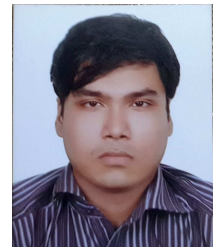

Saifuddin Md. Tareeq, Dr., graduated from Department of Computer Science and Engineering, University of Dhaka and working as Associate Professor at the same Department.

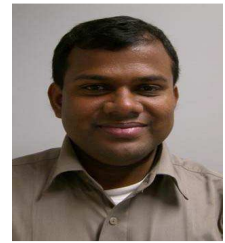